# Teaching and Learning Science with Learning Assistants.

Andrew Ferstl
*Physics Department, Science Education Coordinator, Winona State University*

**Abstract**

The science content classes for elementary education majors at Winona State University used Learning Assistants for the first time during the 2009 – 2010 academic year. Pre-post information was gathered about the Learning Assistants and the pre-service teachers to gauge the effect of this experience on both populations.

**Introduction**

With the need for better PK-16 science teaching and learning, Winona State University developed two interdisciplinary science content courses (Science Education 201 and 203) for elementary education majors during the 2003 – 2004 year. There were several goals of this class. First, we hoped to alleviate some of the trepidation that many Elementary Education majors have about science and in the process increase scientific competency as well as attitudes toward science. Based on attitudinal surveys and scientific reasoning skill tests (the Classroom Test for Scientific Reasoning[i] (CTSR) measures Higher Order Thinking Skills [HOTS]), we are accomplishing this goal. Second, we hoped to model a way to teach science by inquiry that these pre-service teachers can use in their own classrooms. And third, we wanted to impress upon these students that teachers are in many ways scientists themselves as they create lesson plans, implement them, take data on student learning, act on that information to better inform their teaching effectiveness and participate in action research.

Originally the two courses blended biology, chemistry, geoscience, and physics. To implement this class, a team of instructors (one from each of the science disciplines plus a faculty member from Education) co-taught the classes the first time they were offered. Since then the courses have evolved; for example Scie Ed 201 is now a blend of mostly chemistry and physics whereas the Scie Ed 203 is in large part a blend of earth science and life science. But even with these changes we were always able to have at least two science faculty as instructors for the class due to, in part, its interdisciplinary nature as well as the inquiry teaching methods used and the size of the class. With recent events, we've had to rethink the two instructor model. The solution that we tried during the 2009 – 2010 academic year is to continue using two instructors but only one would be a faculty member; the other would be a Learning Assistant (LAs).

LAs[ii] are experienced and talented undergraduates who have been identified by a previous instructor as having excellent interpersonal skills, leadership skills, proficient content knowledge, and a significant potential to be a teacher. Faculty might identify these students, for example, based on the performance of the student in a cooperative group setting or on other student activities.

LAs have been successfully used at other institutes[iii] and can serve as a recruiting tool for future STEM teachers[iv].

**Benefits of Learning Assistants**

Given this information, we felt that Learning Assistants would not only be consistent with the goals of the class (presented in the previous paragraph) but may actually help us better meet them. We proposed the following benefits:

Benefits to the LA.
    (1) Content knowledge gains[v] and scientific reasoning gains.
    (2) More reflective of his/her learning and the learning of others. The LA is better equipped to think about her/his own thinking (metacognition), as well as thinking about his/her students' thinking. They can critically reflect on how they learned/struggled with a concept, how others learn/struggle through that concept, and how they were able to make sense of it.
    (3) More aware of the inquiry learning environment and how their "knowledge" in other courses may just be memorized facts with little or no utility, whereas the knowledge they developed in the inquiry setting is more robust and applicable in less familiar settings.
    (4) Better understanding of group dynamics and interpersonal skills
    (5) A teaching experience that could potentially happen before students are admitted to the College of Education; thus allowing a more informed choice, on the part of the LA, if they want to continue pursuing the field of teaching.
    (6) Come to appreciate that teachers are scientists too as they perform action research
    (7) Looks good on a resume or transcript

Benefits to the SCIE students
    (1) Receive assistance in a timely fashion. With only one instructor in the classroom, when groups encounter difficulty they may have to wait to have a question answered. With LAs, the group receives feedback/guidance when they need it.
    (2) See a model of scientific problem solving. LAs are not lecturers, but when a student encounters difficulty the LA can ask leading questions to assist the student or group. By doing so, the LA is modeling the process of science.
    (3) See a model of a successful student. LAs have more "street credibility" with students than professors, so when an LA tells students that they have to study more in order to be successful, it usually carries more weight than if the professor says it. In fact, the LAs can be much more critical of students' study habits and performance without losing this credibility, because the LA has successfully completed the course so he/she is a perceived "expert" in getting through.

**LA Recruitment and Training**

To achieve the stated goals, we needed to be selective. All students chosen to be LAs were talented undergraduates who had been identified by a previous instructor (who is familiar with the class) as having excellent interpersonal skills, leadership skills, proficient content knowledge, and a significant potential to be a teacher. The students were not necessarily upper classmen especially since we are using this as a recruiting tool for more STEM teaching majors.

An incentive for students to accept this position was a small stipend.

The sources of LAs included:
   (1) Students who have successfully completed the same class for which they will be an LA
   (2) Students in introductory science classes (biology, chemistry, geoscience, and physics) but not necessarily a teaching major.
   (3) Students who have already declared a teaching major in a science, technology, engineering, or math (STEM) field.

To achieve the benefits, the Learning Assistant were expected to:
   (1) Be a facilitator of student learning during the group activities. Not a lecturer or a grader.
   (2) Help in planning and implementing lessons. In other words the LA will attend the weekly coordination meeting led by the instructors and the science education coordinator, discuss student alternative conceptions and teaching techniques to overcome them, provide input on the lesson plans, and help with equipment set up and clean up.
   (3) Participate in LA training consisting of a weekly discussion group with the cohort of other LAs, facilitated by WSU's Science Education Coordinator as well as Winona State University's Tutoring Coordinator.
   (4) Keep a weekly journal/portfolio in which s/he will record/reflect on specific questions/ideas proposed during the weekly discussion group.
   (5) Model the behaviors of a successful pre-service teacher

**Outcomes and Assessment**

To help us guide our study of the impact of LAs on the classroom, we determined what we thought the outcomes would be and how we would assess them. You will notice that the benefits listed above have some overlap here but it is not a complete overlap since we have not yet determined how to measure some of the listed benefits.

| Outcome | Assessment Plan |
|---|---|
| The LA's will have increased their scientific content knowledge and scientific thinking skills | Thinking Skills will be measured by pre-post gains on the Classroom Test for Scientific Reasoning (CTSR)<br><br>Content gains will be measured by assessing weekly reflective journals kept by the LAs |
| The students, who interact with the LAs, | Thinking Skills will be measured by pre-post |

| will increase their scientific content knowledge and scientific thinking skills | gains on the Classroom Test for Scientific Reasoning (CTSR) Content gains will be measured by assessing student performance on exams and homework. |
|---|---|
| The LA will be a more reflective practitioner about aspects of teaching and learning | Measured by the change in the weekly journal/portfolio entries from the beginning of the semester to the end to look for demonstrated growth in their beliefs of teaching and learning (see Rubric in Appendix A for assessment criteria) |
| More students will declare a major in STEM education careers | Keep track of the number of Elementary Education majors that change their major to Middle School Science specialty or a STEM Secondary teaching major. |
| The students, who interact with the LAs, will have an increased positive attitude toward science and science teaching | Measured by pre-post attitudinal survey such as the STEBI[vi] (Student Teaching Efficacy Belief Inventory). |

To be able to compare how two faculty instructors compare to one faculty instructor plus an LA, we will be able to only examine the CTSR gains since this is the only similar data that we have from past classes.

**Results**

Eight Learning Assistants (one for each class) were recruited, trained, and employed during the Fall 2009 – Spring 2010 academic year. They had the following demographics:

- For Fall 2009
    - One was a secondary Social Science/History teaching major – junior, male; Two were elementary education majors with middle school science emphasis – both seniors, female; One was an undeclared major – a sophomore, female
- For Spring 2010
    - One was a secondary Earth Science Teaching Major – a junior, female; One was an elementary education major with a middle school science emphasis – sophomore, male; One was an elementary education major with a math emphasis – junior, female; One was an elementary education major – junior, female

Each LA assisted with a classroom of roughly 30 students.

At the level of the students in the class, the Learning Assistants were a success. This was measured in four ways:

1. First the students were given the CTSR pre and post to measure the change in their ability to reason like a scientist. Scie Ed 201/203 was always taught with two faculty instructors prior to this year and always had a good gain of about 1 to 1.5 points on a 13 points scale (about twice as much as what traditional science classes get). With only one instructor and an LA in the classroom, the gains remained the same. This is an encouraging result.

2. Second, end-of-semester anonymous student evaluations had positive results for the role of the LA.
    a. 88% agreed or strongly agreed that the Learning Assistants were a valuable member of the teaching team.
    b. 92% of the students understood, to a satisfactory degree, the role of the LA and appreciated the helpfulness of him/her.
3. Third, the students were given the STEBI pre and post instruction. 21 of the 23 questions on this survey either saw a positive change or no change. (Questions 10 and 13 saw a negative change.) One example is the question "I know the steps necessary to teach science concepts effectively". Pre only 5% answered "agree or strongly agree"; Post 64% answered "agree or strongly agree".
4. Finally, the students were able to use the LAs to ask the "dumb questions" that they were afraid to ask the instructor. This allowed the LA to guide the students but it also allowed the LA to give feedback to the instructor about where students were stuck.

The Learning Assistants were directly assessed in two ways.

1. First, the CTSR was given pre-post to the LAs. There was not a significant increase in the scores. This, in hindsight, is not surprising since they all scored very high on the CTSR at the beginning (i.e. we purposely recruited "the brightest and best") and thus had little room for improvement.
2. Second, the STEBI was given to the LAs. There was a positive change on 4 of the 23 questions. One of those was "When a student does better than usual in science, it is often because the teacher exerted a little extra effort". More of the LAs agreed or strongly agreed with the statement at the end of the semester.

Another measure of the effect of the Learning Assistants was on their role as a peer mentor for the students. Many of the students asked for help from the LAs outside of class (usually for a study session in preparation for an exam). And the positive role modeling that they performed helped recruit the next set of LAs for Fall 2010.

There seems to also be an increase in the number of students who want to add on a middle school science specialty to their elementary education licensure program. It is difficult to gauge the true effect of this since it is always small numbers. But for this year, we had four students interested in declaring a middle school science specialty after their experience with the class. Two of those are going to be Learning Assistants for the next year.

There were also at least two lessons learned during this experience:

- Lesson one: LAs really enjoyed the experience and it was very eye opening for them to see what happens behind-the-scenes for a class. They found the experience extremely rewarding and used the tutoring skills in other classes.
- Lesson two: The students in the class did rely on the LAs and found them to be a great resource. The students would open up more to the LA and ask the "dumb questions". The LAs then used this type of questioning to inform the instructor about what was going well and what needs to be improved.

**Future**

We wish to continue to use the LAs but the funding source has been depleted. So, to keep the program sustainable, we are relying on (a) the positive influence/experience of our current LAs and (b) a non-monetary reward of a 1 credit course on the LAs transcript showing that they've had this experience. So far, we have recruited all required LAs for the next semester.

**Acknowledgements**

This project was made possible through a Center for Teaching and Learning grant with funding from the Office of the Chancellor, Minnesota State Colleges and Universities.